\documentclass[showpacs,showkeys,preprintnumbers,pre,amsmath,amssymb,superscriptaddress,longbibliography]{revtex4-2}
\pdfoutput=1

\usepackage{graphicx}
\usepackage{geometry}
\usepackage{booktabs}
\usepackage{array}
\usepackage{epic,eepic,amsmath}
\usepackage{graphics}
\usepackage{epsfig}
\usepackage{amsmath}
\usepackage{bm}
\usepackage{amssymb}

\usepackage{color}
\usepackage{amssymb}
\usepackage{amsmath}
\usepackage{graphicx}
\usepackage{colortbl}

\usepackage{amssymb}
\usepackage{enumerate}
\usepackage{amsthm,amsfonts,amsmath}
\usepackage{bm}
\usepackage{graphicx}
\usepackage{graphics,epsf}
\usepackage{color}
\usepackage{xspace}
\usepackage{accents}

\renewcommand{\vec}[1]{\ensuremath{\mathbf{#1}}}

\newcommand*{\vcenteredhbox}[1]{\begingroup       
\setbox0=\hbox{#1}\parbox{\wd0}{\box0}\endgroup}

\begin{document}

\title{Lattice Boltzmann simulations of droplet breakup in confined and time-dependent flows}

\author{Felix Milan}
\email{Electronic address: felix.milan@roma2.infn.it}
\affiliation{Department of Physics and INFN, University of Rome ``Tor Vergata'', Via della Ricerca Scientifica 1, 00133 Rome, Italy}
\affiliation{Department of Applied Physics, Eindhoven University of Technology, Eindhoven 5600 MB, The Netherlands}

\author{Luca Biferale}
\email{Electronic address: biferale@roma2.infn.it}
\affiliation{Department of Physics and INFN, University of Rome ``Tor Vergata'', Via della Ricerca Scientifica 1, 00133 Rome, Italy}

\author{Mauro Sbragaglia}
\email{Electronic address: sbragaglia@roma2.infn.it}
\affiliation{Department of Physics and INFN, University of Rome ``Tor Vergata'', Via della Ricerca Scientifica 1, 00133 Rome, Italy}

\author{Federico Toschi}
\email{Electronic address: f.toschi@tue.nl}
\affiliation{Department of Applied Physics, Eindhoven University of Technology, Eindhoven 5600 MB, The Netherlands}
\affiliation{Department of Mathematics and Computer Science, Eindhoven University of Technology, Eindhoven 5600 MB, The Netherlands}
\affiliation{CNR-IAC, Via dei Taurini 19, 00185 Rome, Italy}



\begin{abstract} We study droplet dynamics and breakup in generic time-dependent flows via a multicomponent lattice Boltzmann algorithm, with emphasis on flow start up conditions. We first study droplet breakup in a confined oscillatory shear flow via two different protocols. In one setup, we start from an initially spherical droplet and turn on the flow abruptly (``shock method''); in the other protocol, we start from an initially spherical droplet as well, but we progressively increase the amplitude of the flow, by allowing the droplet to relax to the steady state for each increase in amplitude, before increasing the flow amplitude again (``relaxation method''). The two protocols are shown to produce substantially different breakup scenarios. The mismatch between these two protocols is also studied for variations in the flow topology, the degree of confinement and the inertia of the fluid. All results point to the fact that under extreme conditions of confinement the relaxation protocols can drive the droplets into metastable states, which break only for very intense flow amplitudes, but their stability is prone to external perturbations, such as an oscillatory driving force.
\end{abstract}

\maketitle

\section{Introduction}
\label{sec:intro} 
Fluid dynamics phenomena, involving droplet dynamics, deformation, and breakup, are prominent in the field of microfluidics and even in general complex flows at larger scales. Beyond the practical importance in a variety of concrete applications~\cite{Flumerfelt72,Singh2010,Shewan2013,Rodriguez2015}, they are also relevant from the theoretical point of view, due to the complexity of the physics involved~\cite{Fortelny2019,Taylor1932,Greco02,GuidoRev,Janssen08}. Droplet deformation is characterized via the capillary number,

\begin{equation}
\label{eq:capillary}
\mbox{Ca}=\frac{\eta_\text{s} R G}{\sigma} \text{,}
\end{equation}

where $\eta_\text{s}$ is the dynamic viscosity of the solvent, $R$ the radius of the initially undeformed spherical droplet, $\sigma$ the surface tension and $G$ the shear rate intensity~\cite{Taylor1932,Greco02b}. The value of $\mbox{Ca}$ at break up is denoted by $\mbox{Ca}_{\text{cr}}$, the critical capillary number. A lot of attention has been dedicated to droplet deformation and breakup in stationary flows~\cite{Fortelny2019,Acrivos1964,Lyngaae1990}, and, in particular, the effect of the degree of confinement on the flow dynamics~\cite{Vananroye08,Janssen10,Guido11}. The degree of confinement is parametrized by the ratio $\alpha = 2 R / L$, where $L$ denotes the shear wall separation. Confinement is frequently encountered in experimental setups of droplet dynamics in simple shear flows~\cite{Vananroye08,Janssen10,Guido11,Bruyn,Cardinaels09,Cardinaels10,Cardinaels11,Cardinaelsetal11b,Vananroye06,Vananroye07,Vananroye11b,VanPuyvelde08} and can be enhanced by changing $\alpha$. There are some theoretical models which were developed to capture the experimental phenomenology of confined droplet dynamics, analytical models~\cite{ShapiraHaber88,ShapiraHaber90}, which extended the theory by Taylor~\cite{Taylor1932,Taylor34}, and phenomenological models~\cite{MaffettoneMinale98,Minale08,Minale10b}. The validity of the analytcial models were verfied in Ref.~\cite{Sibillo06} and the phenomenological models in Ref.~\cite{Minale10}. Of particular interest are the results in Ref.~\cite{Janssen10}, which show that, for non vanishing $\alpha$ breakup differs substantially from the unconfined shear case both qualitatively and quantitatively for all viscosity ratios $\chi=\eta_{\tiny \mbox{d}} / \eta_{\tiny \mbox{s}}$, where $\eta_{{\tiny \mbox{d}},{\tiny \mbox{s}}}$ is the dynamic viscosity of the droplet (d) or solvent (s) phase. Additionally,  the dependency of the critical capillary number $\mbox{Ca}_{\text{cr}}$ on the droplet's inertia is a central area of interest~\cite{GuidoRev,Alguacil2004,Renardy2001,Caserta07,Chaffey,Chinyoka05,Elmendorp,Feigl02,Grace,Komrakova13,Li2000,Li08,Liu,Mighri,Mighri06,Mukherjee09,PillapakamSingh04,AggarwalSarkar07,AggarwalSarkar08,Sibillo04,TuckerMoldenaers02,Zhao07}, with the degree of inertia being given by the Reynolds number,

\begin{equation}
\label{eq:reynolds}
\mbox{Re} = \frac{G R^2}{\nu_{\tiny \mbox{s}}} \text{,}
\end{equation}

where $\nu_{\tiny \mbox{s}}$ is the kinematic viscosity of the solvent. Furthermore, breakup is influenced by the start up conditions, as demonstrated in experimental and theoretical studies~\cite{Torza72,Hinch80,Brady82,Bentley86,Renardy08}. This phenomenon is rather subtle and can have different effects depending on the protocol in use. The dependency on the rate of increase of the shear rate $G$ was confirmed by~\cite{Torza72} via supporting calculations based on the model by Taylor~\cite{Taylor1932}. A theoretical model developed by Hinch \emph{et al.}~\cite{Hinch80} shows that stable droplet equilibria below the critical capillary number $\mbox{Ca}_{\text{cr}}$ breakup are only possible for a sufficiently low increase in $G$. Furthermore, Renardy~\cite{Renardy08} has shown that although these stable equilibria require a slow increase in the shear rate $G$ they are unique and do not depend on the rate of change of $G$. We stress that even though the effect of start up conditions on break up has been investigated~\cite{Torza72,Hinch80,Brady82,Bentley86,Renardy08}, the role of confinement with varying start up conditions on droplet dynamics and break up is not clear. Moreover, it is unclear how break up is affected, if the flows are time-dependent~\cite{Cox69,Farutin2012,Cavallo02,Yu02,Milan2018}. The aim of the present paper is to take a step further in this direction. With the use of numerical simulations we show that at capillary numbers close to breakup, confinement allows for the existence of a metastable flow configuration next to the solution of the Stokes equation found in Ref.~\cite{Renardy08}. This metastable state is prone to perturbations and collapses to the Stokes solution, if we have a time-dependent flow with a sufficiently large shear frequency. It should be stressed that this result is unique to the case of a confined droplet in an oscillatory shear, as this metastable configuration is not present neither for an unconfined droplet in an oscillatory shear flow nor in the case of an oscillatory elongational flow. Our studies can be seen as an extension to Refs.~\cite{Renardy2001,Renardy08}, where the influence of inertia on droplet breakup was studied, whereas we deal with time-dependent cases, where the temporal rate of change of the shear intensity is comparable to the droplet relaxation time,

\begin{equation}
\label{eq:drop_time}
t_d = \frac{\eta_{\tiny \mbox{d}} R}{\sigma} \text{.}
\end{equation}

This work is a follow up study of~\cite{Milan2018}, where stable time-dependent droplet dynamics was investigated via a multicomponent lattice Boltzmann scheme and a phenomenological model~\cite{Minale08,MaffettoneMinale98}. It was found that droplet deformation depended strongly on an external timescale, the oscillation frequency of an oscillatory shear flow, for a confined droplet. For relatively large oscillation periods close to the value of $t_d$ the droplet is hardly deformed by the solvent shear flow, which was described as the ``transparency effect" in Ref.~\cite{Milan2018}. The findings in Ref.~\cite{Milan2018} have been validated by comparing the lattice Boltzmann results to the results obtained via a phenomenological droplet deformation model, the Maffettone-Minale model~\cite{Minale08,MaffettoneMinale98}. \\

\noindent This paper is organized as follows: Sec.~\ref{sec:latticeboltzmann} gives a brief overview on the lattice Boltzmann algorithms and models in use. In Sec.~\ref{sec:methodology}, we outline the general details of droplet break up with an emphasis on confined systems and simple shear flows. In Sec.~\ref{sec:simple_shear}, we investigate break up in a time-dependent (oscillatory) shear flow under strong confinement. A mismatch between two protocols, involving different start up conditions of the flow, leads us to investigate break up conditions under the influence of inertia (Sec.~\ref{sec:linear}) and the effect of confinement (Sec.~\ref{sec:confinement}) Moreover, we check whether the protocol mismatch depends on the flow topology (Sec.~\ref{sec:elongational}).

\section{lattice Boltzmann algorithms and methods}
\label{sec:latticeboltzmann}

We use lattice Boltzmann simulations~\cite{Benzi92,Succi01} to study droplet break up in confined and time-dependent shear and elongational flows. The lattice Boltzmann method (LBM) has been extensively used in the field of microfluidics, including extensions to accommodate nonideal effects~\cite{Sbragaglia12}, coupling with polymer micro-mechanics~\cite{Onishi1} and thermal fluctuations \cite{Varnik11,Xue2018}. LBM has also been used widely for the modeling of droplet break up behavior~\cite{Komrakova13,Liu,Farokhirad,Onishi2,Xi99,Yoshino08,GuptaSbragaglia2014,GuptaSbragagliaScagliarini2015,Chiappini2018,Chiappini2019}. 
To model multicomponent systems with the lattice Boltzmann Model (LBM) we need to account for interfacial forces between different fluid components. This can be achieved with the Shan-Chen multicomponent model (SCMC) \cite{Shan93,Shan94}, a diffuse interface model in the framework of the LBM. The hydrodynamical quantities, mass and momentum densities, can then be described as:

\begin{align}
\label{eq:multi_momentum}
\rho(\vec{x}, t) & = \sum_{\sigma} \sum_i g_i^{\sigma}(\vec{x}, t) \text{,} \notag \\
\rho(\vec{x}, t) \vec{u}(\vec{x}, t) & = \sum_{\sigma} \sum_i g_i^{\sigma}(\vec{x}, t) \vec{c}_i \text{,}
\end{align}

where $g_i^{\sigma}(\vec{x},t)$ denotes the populations in the LBM model for the fluid component $\sigma$ and $\vec{c}_i$ are the lattice velocities. For example, for a two component system with species $A$ and $B$ the index $\sigma$ can take the values $\sigma = A$ and $\sigma = B$. The interaction at the fluid-fluid interface \cite{Sbragaglia2013,Sega2013} is given by:

\begin{equation}
\label{eq:scmc_interface}
\vec{F}^{\sigma}(\vec{x},t) = - \rho_{\sigma}(\vec{x},t) \sum_{\sigma' \neq \sigma} \sum_{i=1}^{N} \mathcal{G}_{\sigma, \sigma'} w_i \rho_{\sigma'}(\vec{x,t} + \vec{c}_i) \vec{c}_i \text{,}
\end{equation}

where $\rho_{\sigma}(\vec{x},t)$ is the density field of the fluid component denoted by $\sigma$. $\mathcal{G}_{\sigma,\sigma'}$ is a coupling constant for the two phases $\sigma$ and $\sigma'$ at position $\vec{x}$ and $w_i$ are the lattice isotropy weights. We use the same open flow boundary conditions as outlined in Ref.~\cite{Milan2018}. To use arbitrary boundary values of the density $\rho(\vec{x}, t)$ and velocity $\vec{u}(\vec{x},t)$ fields of the solvent fluid we use ghost populations (or halos), which store the equilibrium distribution functions $g_i^{\text{eq}}$ of the boundary density and velocity fields. The equilibrium distribution functions $g_i^{\text{eq}}$ are given by:

\begin{equation}
\label{eq:equilibrium}
g_i^{\text{eq}} (\vec{x},t) = \rho_b(\vec{x},t) w_i \left ( 1 + 3 \, \vec{c}_i \cdot \vec{u} + \frac{9}{2} (\vec{c}_i \cdot \vec{u})^2 - \frac{3}{2} \vec{u}^2 \right ) \text{,}
\end{equation}
 
with $w_i$ being the lattice weights for the set of lattice vectors $\vec{c}_i$, and $\rho_b(\vec{x},t)$ the density field at the simulations domain boundary. Thus the ghost distributions update the boundary nodes during the LBM streaming step and effectively simulate an open flow boundary given by the chosen density $\rho_b(\vec{x}, t)$ and velocity $\vec{u}(\vec{x},t)$ fields of the outer fluid~\cite{Milan2018}. The streaming and collision steps are given by the lattice Boltzmann equation:

\begin{equation}
\label{eq:lbe}
g_i(\vec{x} + \vec{c}_i \Delta t, t + \Delta t) - g_i(\vec{x}, t) = \Omega(\{g_i(\vec{x},t)\}) \text{,}
\end{equation}

where $\Omega(\{g_i(\vec{x},t)\})$ is the collision operator depending on the whole (local) set of lattice populations and $\Delta t$ is the simulation time step. For MRT (multi-relaxation timescale) the collision operator is linear and contains several relaxation times linked to its relaxation modes (depending on the lattice stencil) \cite{Humieres02}. One relaxation time $\tau$ is directly linked to the kinematic viscosity $\nu$ in the system

\begin{equation}
\label{eq:viscosity}
\nu = \frac{1}{3} \left ( \tau - \frac{1}{2} \right ) \text{,}
\end{equation}

which is one of the primary links between the LBM scheme and hydrodynamics \cite{Benzi92,Succi01}. The boundary scheme described here is not strictly mass conserving, so we correct the local population mass densities To cure mass conservation~\cite{Mattila09,Hecht10,Zou97}. This is not the case in unconfined system, where we can accept small mass fluctuations of both droplet and solvent, but have to reinject mass into the droplet~\cite{Biferale11}.

\section{Simulation setup and definitions}
\label{sec:methodology}

In this section we define what we mean when we speak of droplet break up and characterize the simulation setups. We deal with both a confined droplet in a simple shear flow and an unconfined droplet in a uniaxial extensional (elongational) flow. The velocity gradient matrix for both shear and elongational flows is given by

\begin{equation}
\label{eq:shear_matrix}
\nabla \vec{v} = \frac{G}{2}
\begin{pmatrix}
\beta && 0 && 2 (1 - \beta) \\
0 && \beta && 0 \\
0 && 0 && -2 \beta \\
\end{pmatrix} \text{,}
\end{equation}

where $\lVert \nabla \vec{v} \rVert = G$ and $\beta$ is a parameter characterising the flow type. The shear flow setup is equivalent to the one used in Ref.~\cite{Milan2018} with $\beta=0$ in Eq.~(\ref{eq:shear_matrix}) except that the flow is unconfined and elongational with an oscillatory velocity gradient amplitude $G(t)$ given by Eq.~(\ref{eq:shear_matrix}) with $\beta = 1$. Droplet deformation can be characterized by the capillary number $\mbox{Ca}$. In the case of a shear flow including confinement the shear rate is given by

\begin{equation}
\label{eq:shear}
G = \frac{2 u_0}{L_z} \text{,}
\end{equation}

with $L_z$ being the channel width responsible for the droplet confinement and $u_0$ being the maximum wall velocity amplitude. This definition may also be extended to time-dependent shear flows~\cite{Milan2018}

\begin{equation}
\label{eq:shear_time}
G(t) = \frac{2 u(t)}{L_z} \text{.}
\end{equation}

\begin{figure}[!htbp]
\centering
\includegraphics[scale=0.55]{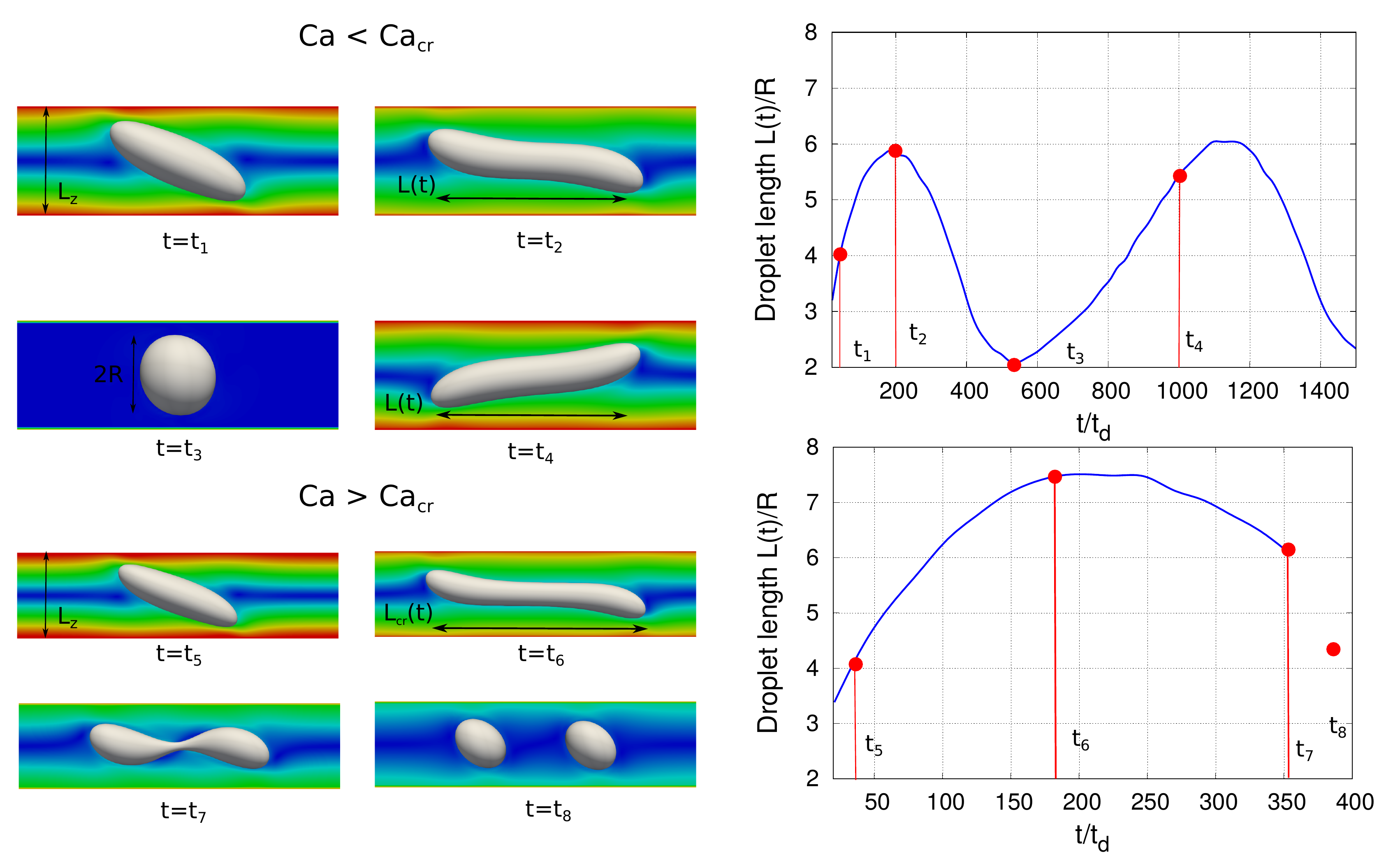}
\caption{Snapshots of a droplet in a confined oscillatory shear flow with a nondimensionalized oscillation frequency $\omega_f t_d$. Snapshots of the droplet in the velocity field are shown for $\mbox{Ca} < \mbox{Ca}_\text{cr}$ and $\mbox{Ca} > \mbox{Ca}_\text{cr}$. The plots on the right panel show the time evolution of the normalized droplet length $L(t)/R$. The degree of confinement of the system is given by $\alpha = 2 R / L_z$, where $R$ is the droplet radius of the undeformed droplet and $L_z$ is the wall separation.}
\label{fig:droplet_sketch}
\end{figure}

In accordance with~\cite{Janssen10} we define the critical capillary number $\mbox{Ca}_\text{cr}$ as the value of $\mbox{Ca}$ for which an initially spherical droplet breaks up, which is achieved by a sudden increase in the shear rate amplitude $G$. We refer to this break up protocol as the \emph{Shock Method}. In addition we can gradually increase the shear rate $G$ starting from a value for which the droplet is only marginally deformed~\cite{Torza72,Hinch80,Renardy08}. A fixed increase $\Delta G$ (or $\Delta u_0$ in the case of Eq.~(\ref{eq:shear_time})) is equivalent to a fixed increment rate $\Delta \mbox{Ca}$ for the capillary number. This way the droplet and the solvent flow are given more time to relax to their respective equilibrium distributions at specific $\mbox{Ca}$. We call this protocol the \emph{Relaxation method}. It should be stressed that breakup in the relaxation method has a small dependency on $\Delta \mbox{Ca}$. If $\Delta \mbox{Ca}$ is very large, e.g. $\Delta \mbox{Ca} \sim \mbox{Ca}_\text{cr}$, the value for $\mbox{Ca}_\text{cr}$ will be the same as the one obtained through the shock method. Thus, the $\Delta \mbox{Ca}$ has to be chosen sufficiently small enough for the relaxation method to work. Essentially, the relaxation method captures  the deformation history of the droplet before breakup with an accuracy given by $\Delta \mbox{Ca}$ contrary to the shock method. The relaxation method is especially important for droplet dynamics in palatially evolving shear flows in the case of a smoothly varying local shear both spatially and temporarily. A variation of the relaxation method for time-dependent oscillatory flows, i.e. where the shear amplitude $G(t) = G_0 \cos(\omega t)$, is to consider the flow and droplet configuration at a capillary number $\mbox{Ca}$ close to $\mbox{Ca}_\text{cr}$ and then to increase the oscillatory shear frequency $\omega_f = \omega / (2 \pi)$ until break up, starting from the stationary case of $\omega_f = 0$. As in Ref.~\cite{Milan2018} we use a dimensionless frequency $\omega_f t_d$ in our discussion, where $t_d$ is the droplet relaxation time defined in Eq.~(\ref{eq:drop_time}). In the presence of a flow with nonzero frequency $\omega_f t_d$, we focus on $\mbox{Ca}_{\text{max}}$, which denotes the maximum value of the time-dependent capillary number $\mbox{Ca}(t)$ over one oscillatory cycle~\cite{Milan2018}. An instance of droplet break up in an oscillatory simple shear flow is depicted in Fig.~\ref{fig:droplet_sketch}. The droplet is oscillating between two maximally elongated states for $\mbox{Ca} < \mbox{Ca}_\text{cr}$ and breaks up during the flow build up for $\mbox{Ca} > \mbox{Ca}_\text{cr}$ in the case of the shock method. The droplet elongation is characterized by the droplet length $L(t)$, which is defined as the longest axis of the elongated droplet, and $L_{\text{cr}}$ denotes the droplet length in the critical case $\mbox{Ca} \geq \mbox{Ca}_\text{cr}$. The time evolution of $L(t)$ is also shown for the two cases $\mbox{Ca} < \mbox{Ca}_\text{cr}$ and $\mbox{Ca} > \mbox{Ca}_\text{cr}$ in Fig.~\ref{fig:droplet_sketch}, which shows that break up occurs at around $t = 17000$ lbu with lbu denoting lattice Boltzmann Units. In all simulations in this article the viscous ratio $\chi \equiv 1$ and the density ratio $\rho_{\tiny \mbox{d}} / \rho_{\tiny \mbox{s}} \equiv 1$. If not explicitly stated otherwise, then the confinement ratio for simple shear flows $\alpha \equiv 2 R / L_z$, where $R$ is the radius of the spherical undeformed droplet and $L_z$ the channel width, is set to $\alpha = 0.75$.

\section{Droplet break up in an oscillatory shear flow}
\label{sec:simple_shear}

\begin{figure}[!htbp]
\centering
\vcenteredhbox{\includegraphics[scale=0.1, trim={10mm, 100mm, 10mm, 100mm}, clip]{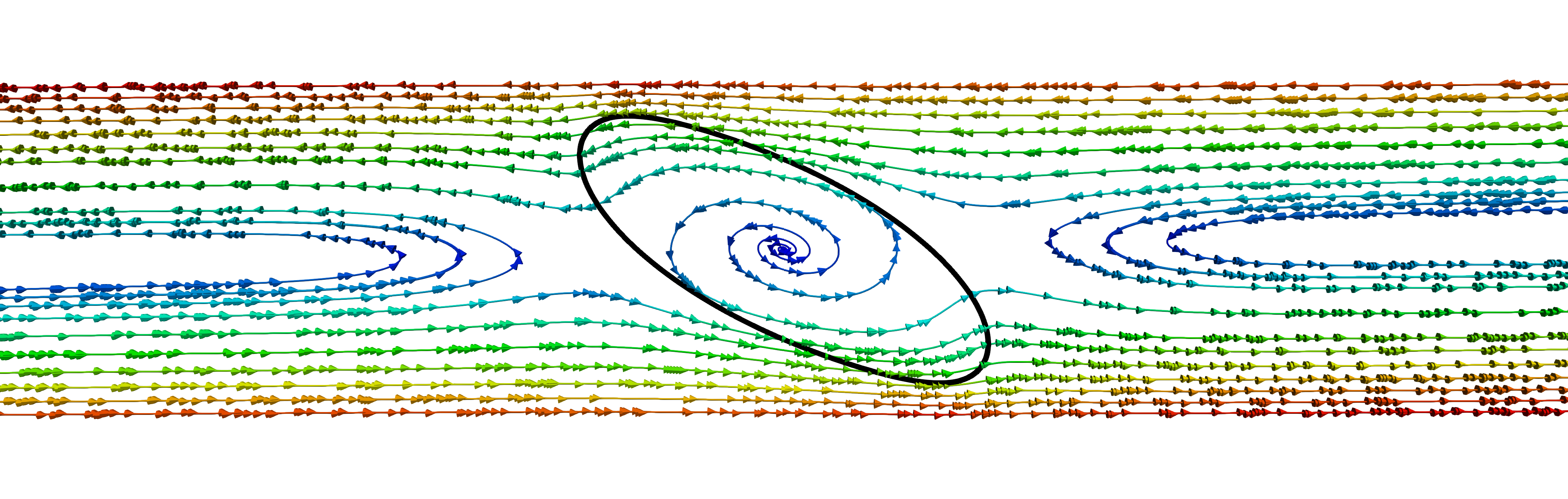}}
\vcenteredhbox{\includegraphics[scale=0.2, trim={10mm, 100mm, 10mm, 100mm}, clip]{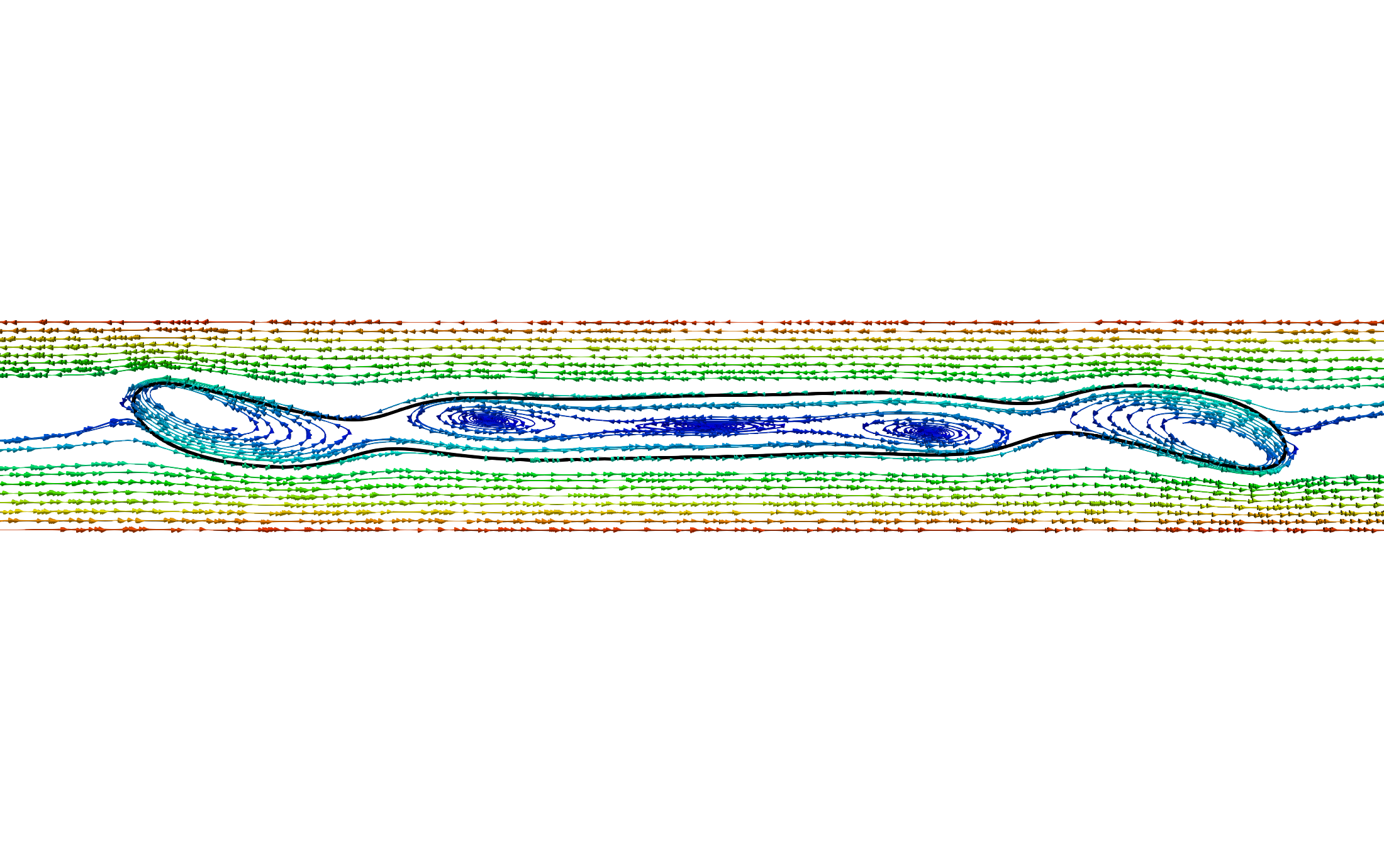}}
\caption{Planar cut of a droplet in a shear flow, featuring an ellipsoidally deformed droplet and large droplet deformation before breakup. The droplet contours are shown in black and the velocity field is visualized by streamlines coloured according to the velocity magnitude.}
\label{fig:droplet_shear}
\end{figure}

\begin{figure}[!htbp]
\centering
\includegraphics[scale=0.73]{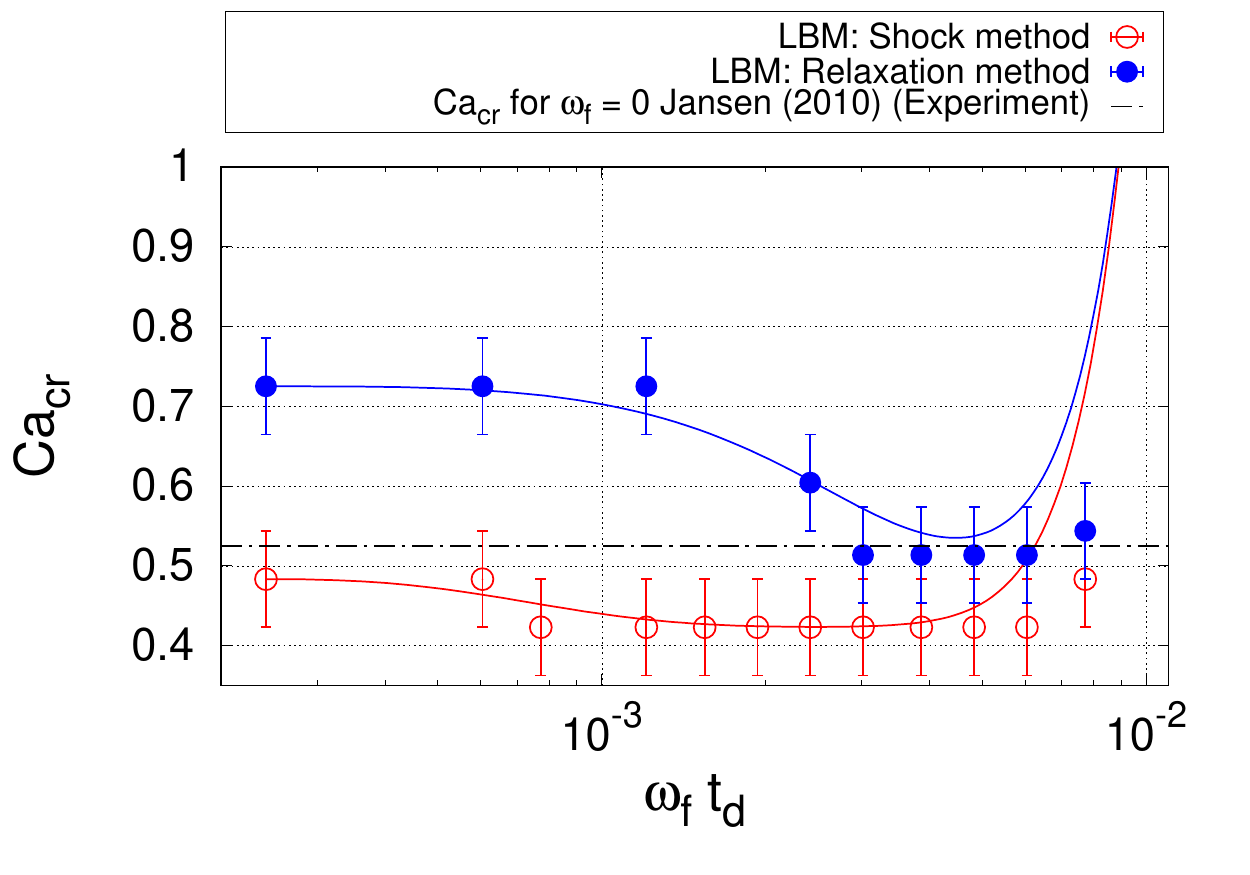}
\caption{Critical capillary number $\mbox{Ca}_{\text{cr}}$ at varying frequencies $\omega_f t_d$. There is a mismatch between the predictions of the two break up protocols. Whereas droplet break up is largely independent in the case of the \emph{shock method}, except for the asymptotic behavior in the high-frequency region, the \emph{relaxation method} in the low-frequency limit predicts a higher $\mbox{Ca}_{\text{cr}}$ than the ones of the shock method. This mismatch is investigated in the article. The error bars are estimated via steps in the critical capillary number $\Delta \mbox{Ca}$. Both curves are interpolated via bezier curves.}
\label{fig:mm_critical}
\end{figure}

Similarly to Ref.~\cite{Milan2018} we consider a droplet in a confined oscillatory simple shear flow; see Fig.~\ref{fig:droplet_shear}. The setup is shown in Fig.~\ref{fig:droplet_sketch} with a confinement ratio $\alpha = 0.75$ and a time-dependent shear rate $G(t) = 2 u_0 / L_z \cos(2 \pi \omega_f t)$, where $\omega_f$ is the frequency of the outer oscillatory flow~\cite{Cox69,Farutin2012,Cavallo02,Yu02,Milan2018}. Our main focus is the dependency of $\mbox{Ca}_\text{cr}$ on the normalized shear frequency $\omega_f t_d$ of the oscillatory outer flow. Droplet dynamics in oscillating flows may feature a so called transparency effect~\cite{Milan2018}, which states that the droplet is hardly deformed if $\omega_f t_d \sim 0.1$, i.e. the timescale of the oscillating shear flow $1/\omega_f$ is of the similar order as the droplet relaxation timescale $t_d$. The droplet dynamics are hardly influenced by the shear frequency for $\omega_f t_d \sim 10^{-4}$ and the transparency effect is noticeable for $\omega_f t_d \sim 10^{-2}$ and higher frequencies, which leads to a sudden increase in the critical capillary number. To stay in tune with experimental results~\cite{Janssen10,Vananroye06,Vananroye07,Vananroye08}, we limit the range of the critical capillary number close to $\mbox{Ca}_\text{cr} \sim 1.0$. In Fig.~\ref{fig:mm_critical} we can see that the droplet break up behavior is significantly different for our two LBM simulation protocols, the shock and relaxation method. The shock method implies that droplet break up is independent of the oscillatory shear frequency $\omega_f t_d$, significant changes in $\mbox{Ca}_\text{cr}$ only occur close to the transparency effect region at high frequencies ($\omega_f t_d \sim 10^{-2}$). The relaxation method is of a different nature: first of all $\mbox{Ca}_\text{cr}$ in the low-frequency region ($\omega_f t_d \sim 10^{-4}$) is larger than the values obtained with shock method (see also Sec.~\ref{sec:confinement}). Moreover, for intermediate frequencies $\omega_f t_d \sim 5 \times 10^{-3}$ we observe that break up occurs at a significantly smaller $\mbox{Ca}_\text{cr}$ than in the low-frequency range and is now of a comparable value to $\mbox{Ca}_\text{cr}$ obtained via the shock method. The mismatch between the two protocols in the low-frequency regime in Fig.~\ref{fig:mm_critical} is in disagreement with previous studies of start up conditions of droplet break up in confined simple shear flows~\cite{Janssen10,Renardy08}. However, the shock method produces results in accordance with the literature~\cite{Janssen10}, as the dashed line in Fig.~\ref{fig:mm_critical} indicates. It should also be noted, that the destabilization of the ``relaxation branch'' is rather sudden and takes place at very small $\omega_f t_d$. This suggests that the protocol mismatch is due to metastable solution (relaxation method) existing next to a stable solution (shock method) in the low-frequency range $\omega_f t_d \leq 0.02$. The protocol mismatch seems rather puzzling: according to Renardy~\cite{Renardy08} the solution should be unique. However, our setup differs in a few points from the one in Renardy~\cite{Renardy08}. First of all, the droplet is strongly confined ($\alpha=0.75$) in our setup (see Fig.~\ref{fig:droplet_sketch}), which could have a strong effect on the values $\mbox{Ca}_\text{cr}$ for varying start up conditions. Moreover, inertia might stabilize the droplet in the case of the relaxation method. Therefore, the protocol mismatch might disappear in the Stokes limit. In addition, one may also wonder what is the effect of flow topology, as an inherently different flow field might lead to a similar protocol mismatch. Given these considerations, in the following sections, we will investigate the cause of the mismatch by considering both inertial effects, as is the case in Ref.~\cite{Renardy08}, (see Sec.~\ref{sec:linear}) and the importance of confinement in stationary shear flows (see Sec.~\ref{sec:confinement}). Regarding the importance of flow topology, we investigate time-dependent break up in an elongational flow in Sec.~\ref{sec:elongational}.

\section{Inertial effects}
\label{sec:linear}

\begin{figure}[!htbp]
\centering
\includegraphics[scale=0.73]{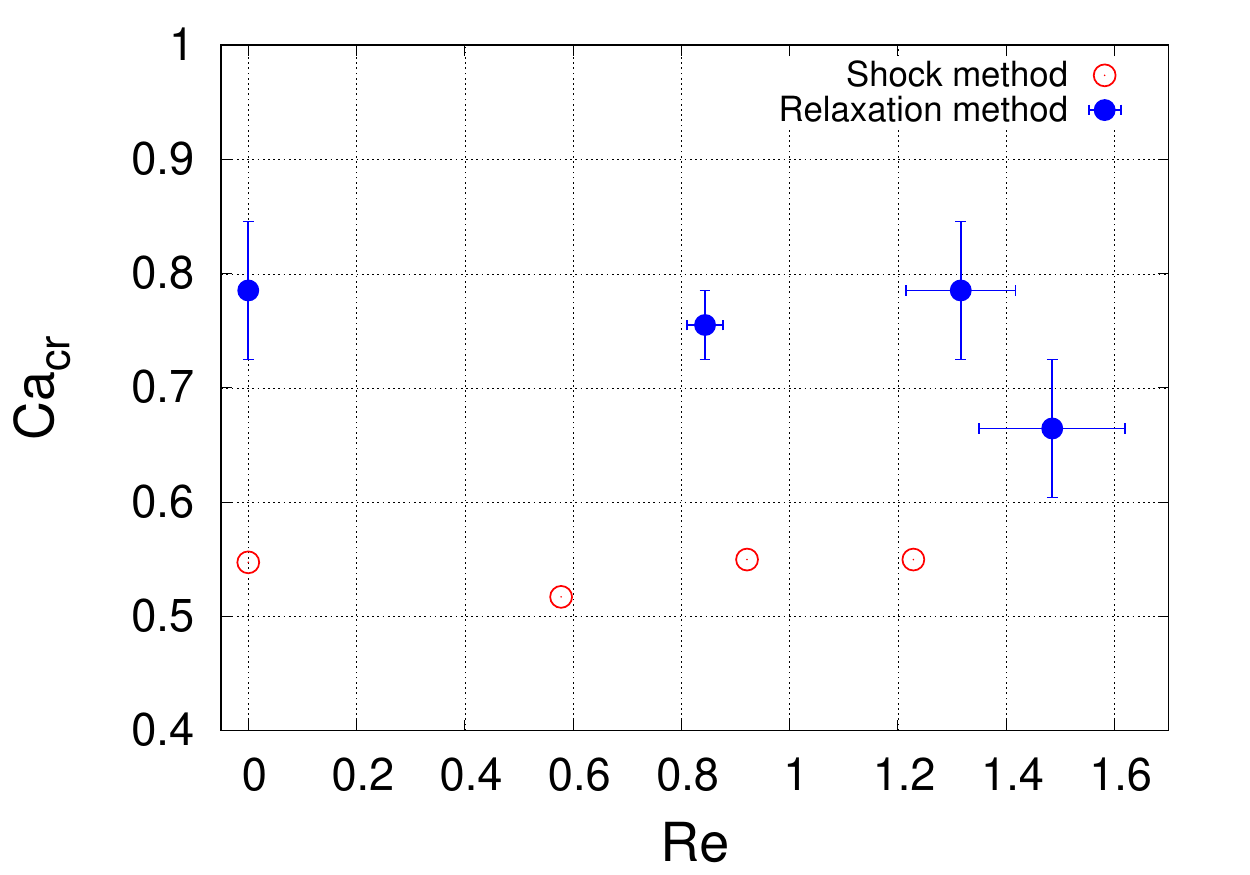}
\caption{$\mbox{Ca}_\text{cr}$ vs Reynolds number $\mbox{Re}$. The mismatch between the shock and relaxation break up protocols does not depend on inertia. This is especially clear in the case of the Stokes solution, for which $\mbox{Re} = 0$. The error bars are estimated via steps in the critical capillary $\Delta \mbox{Ca}$ and Reynolds number $\Delta \mbox{Re}$.}
\label{fig:re_critical}
\end{figure}

In Ref.~\cite{Renardy08} it is shown that the solution of the Stokes equation in confined simple shear flows is unique and does not depend on neither the initial conditions of the droplet nor the solvent flow configuration. Thus, one might think that the protocol mismatch might be due to inertial effects and would disappear, if we were close to the time-dependent Stokes limit of $\mbox{Re} \equiv 0$. Interestingly, the LBM formalism allows us to directly set $\mbox{Re} = 0$, as we can eliminate the nonlinear terms in the equilibrium distribution functions in the LBM algorithm, Eq.~(\ref{eq:equilibrium}), which leads us to a modified equation(~\ref{eq:equilibrium_lin}), accounting only for the linear terms in the velocity field $\vec{u}(\vec{x},t)$. It should be remarked that only the nonlinearites of the Navier-Stokes equation are removed in this way, since the inertia embedded in the time derivative of the velocity field $\vec{u}(\vec{x},t)$ does not disappear and may still play a role during the non steady break-up process. Inertial effects tend to stabilize the droplet~\cite{Brady82,Bentley86} for low $\mbox{Re} < 1$, whereas $\mbox{Ca}_\text{cr} \sim 1 / \mbox{Re}$ for large $\mbox{Re} > 10$~\cite{Renardy2001}. This suggests, that the stabilization effect of low $\mbox{Re}$ are responsible for the protocol mismatch, which consequently should disappear in the Stokes limit $\mbox{Re} = 0$. We investigate the dependency of $\mbox{Ca}_\text{cr}$ on $\mbox{Re}$, as shown in Fig.~\ref{fig:re_critical}. For the case $\mbox{Re} = 0$ we use only the linear terms of the equilibrium distribution functions given by

\begin{equation}
\label{eq:equilibrium_lin}
g_i^{\text{eq,lin}} (\vec{x},t) = \rho_b(\vec{x},t) w_i \left ( 1 + 3 \, \vec{c}_i \cdot \vec{u} \right ) \text{.}
\end{equation}

The simulations are carried out for a stationary shear flow, with the setup described in Fig.~\ref{fig:droplet_sketch}. We can see that the mismatch between the break up protocols, does not depend on inertia and is even present in the Stokes limit of $\mbox{Re}=0$. We conclude that the mismatch between the two break up protocols is not influenced by any stabilization effects of inertia~\cite{Brady82,Bentley86} for the given range of Reynolds numbers $\mbox{Re} \sim 0.0, \ldots, 1.5$.

\section{Confinement effects}
\label{sec:confinement}

\begin{figure}[!htbp]
\centering
\includegraphics[scale=0.73]{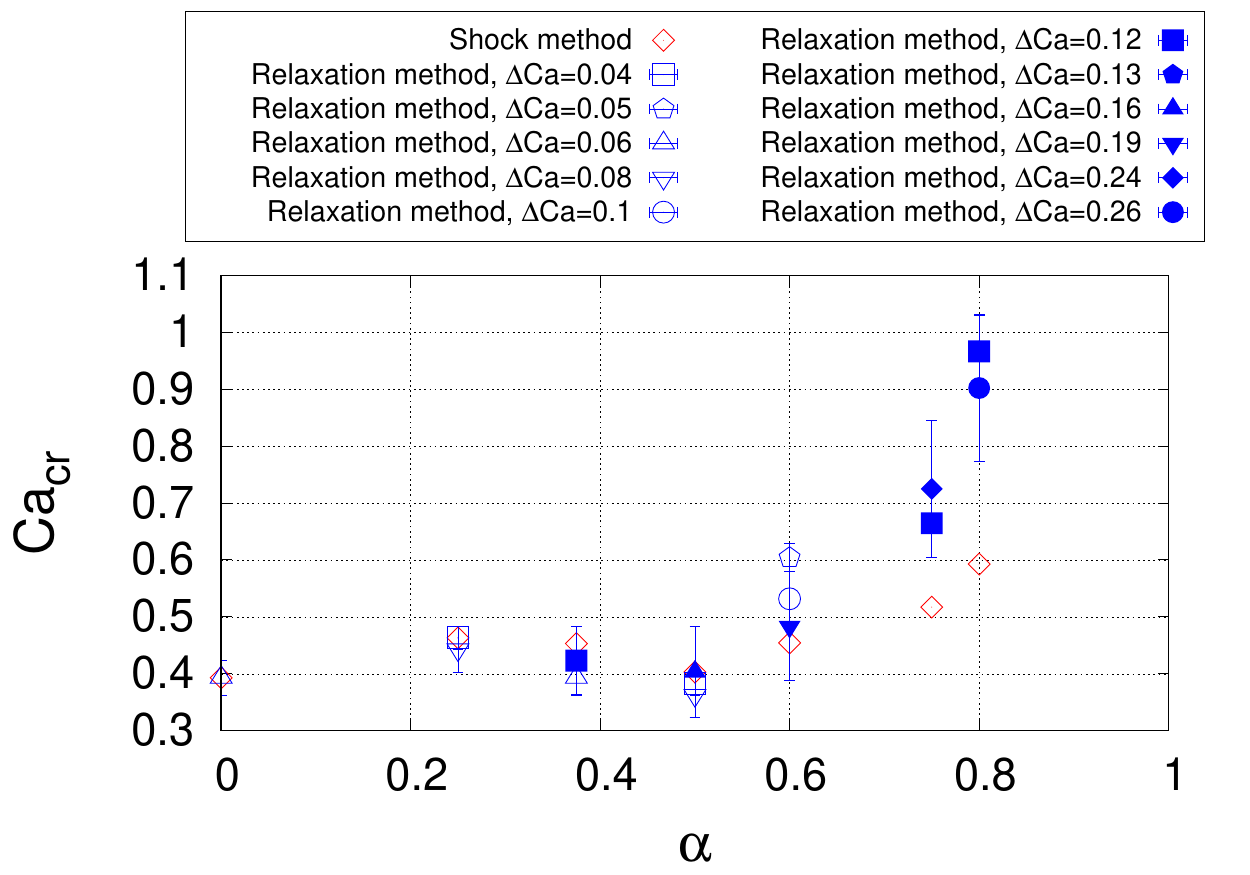}
\caption{Critical capillary number $\mbox{Ca}_\text{cr}$ for different confinement ratios $\alpha = 2 R / L_z$. We compare the values obtained by the LB simulations with the shock method and the ones obtained by the relaxation method. Since the relaxation method is dependent on the start up conditions of the outer flow and the droplet, we provide a range of different increments $\Delta \mbox{Ca}$, where smaller $\Delta \mbox{Ca}$ denote a slower and flow build up and vice versa. The error bars are estimated via steps in the critical capillary number $\Delta \mbox{Ca}$. For each simulation run of the relaxation method with a given $\mbox{Ca}$ we gave the droplet a sufficiently long time to relax to its stationary state.}
\label{fig:aspect_critical}
\end{figure}

\begin{figure}[!htbp]
\centering
\includegraphics[scale=0.73]{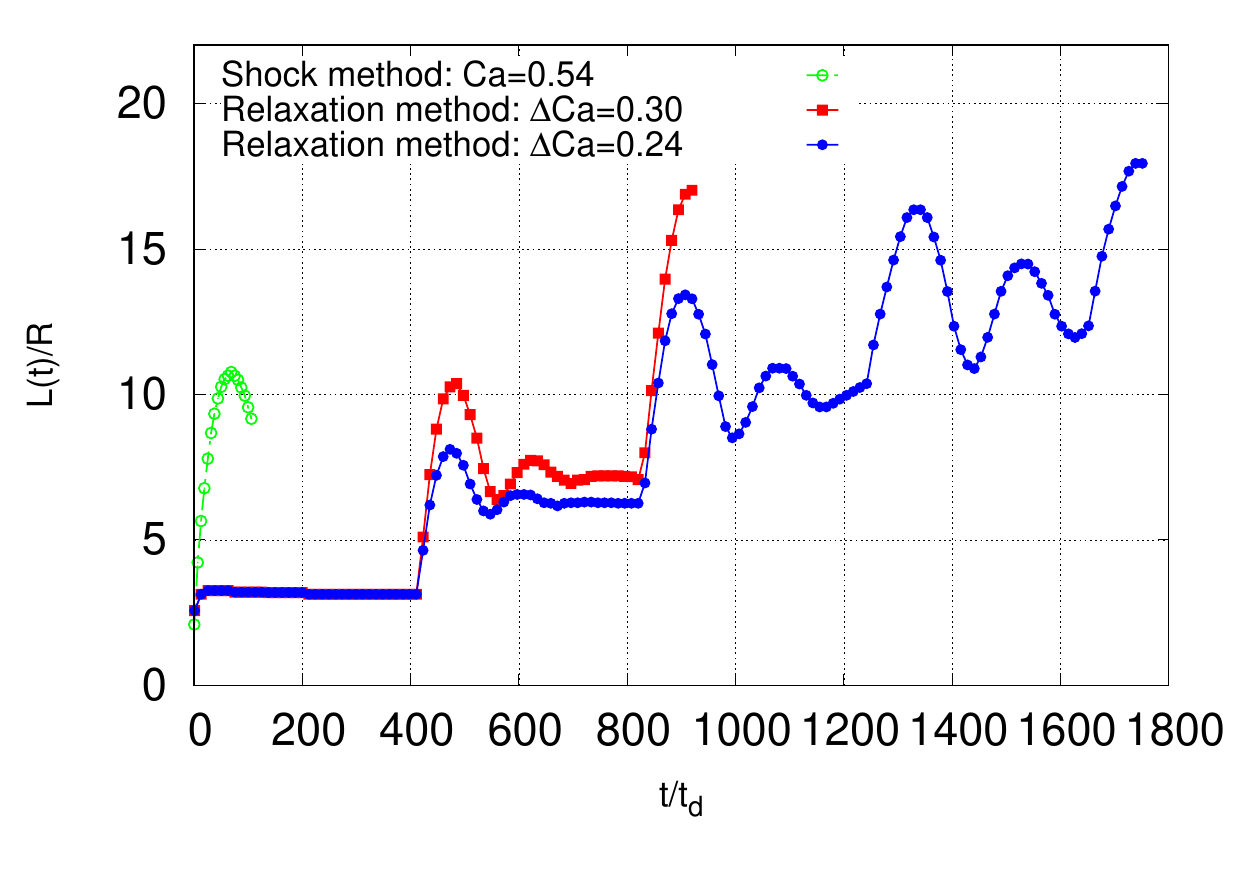}
\caption{Normalized droplet major axis $L(t)/R$ against simulation time $t$ given in units of the droplet relaxation time $t_d$. The droplet breaks up shortly after its maximum elongation for the shock method. Break up in the relaxation method is dependent on the shear rate and thus capillary number increase: (a) for a rate with increment $\Delta \mbox{Ca} = 0.30$ the droplet relaxes after reaching its maximum elongation for the first time to break up at a longer length at a higher $\mbox{Ca}_\text{cr}$ later on; (b) for a smaller capillary number increase $\Delta \mbox{Ca} = 0.24$ the droplet length at $\mbox{Ca}_\text{cr}$ increases even further and the $L(t)$ contains more full extensions and subsequent retractions.}
\label{fig:droplet_length}
\end{figure}

We now focus on both confinement and start up conditions in the shear rate amplitude $G$ for droplet break up in a  stationary shear flow. The setup is once again the one in Fig.~\ref{fig:droplet_sketch}, a confined droplet in a stationary ($\omega_f t_d = 0$) shear flow, but now we vary the confinement ratio $\alpha$ and, in the case of the relaxation method, the rate of change of the shear amplitude $G$, resulting in increments of the capillary number $\Delta \mbox{Ca}$. Our results are summarized in Fig.~\ref{fig:aspect_critical}. We can see, as was shown in Ref.~\cite{Renardy08}, that the critical capillary number $\mbox{Ca}_\text{cr}$ is independent of the start up conditions for low confinement ratios ($\alpha \leq 0.5$), as both the shock method and the relaxation method yield the same results with respect to the simulation errors. However, if the droplet is strongly confined ($\alpha \geq 0.6$), then the two methods yield very different results, with the $\mbox{Ca}_\text{cr}$ predicted by the relaxation method being substantially larger than the one predicted by the shock method. It should be noted, that $\mbox{Ca}_\text{cr}$ is independent of $\Delta \mbox{Ca}$, given that $\Delta \mbox{Ca}$ is small enough, which can be seen from Fig.~\ref{fig:aspect_critical}, where the values of $\mbox{Ca}_\text{cr}$ overlap in respect to their error ranges for different $\Delta \mbox{Ca}$ and the same $\alpha$. Fig.~\ref{fig:droplet_length} shows the length of the elongated droplet as a function of the LB simulation time for the different shear start up methods: we can see that for the shock method droplet break up occurs soon after the maximal elongation, whereas for the relaxation the droplet experiences a sequence of maximal extensions and subsequent retractions after breaking up for a given $\mbox{Ca}_\text{cr}$ at its critical length $L_\text{cr}(t)$. We conclude that both a slow start up of the outer flow (relaxation method) and a strong confinement of the droplet ($\alpha \geq 0.6$) are necessary for the mismatch reported in Fig.~\ref{fig:mm_critical} in the low-frequency limit. The eventual collapse of the relaxation method solution on to the one found by the shock method suggests, that the relaxation method branch in the low-frequency limit in Fig.~\ref{fig:mm_critical} is a metastable state, explaining the high susceptibility to small perturbations and the collapses to the configuration obtained by the shock method for intermediate oscillatory frequencies $\omega_f t_d$.

\section{Flow topology}
\label{sec:elongational} 

\begin{figure}[!htbp]
\centering
\vcenteredhbox{\includegraphics[scale=0.25]{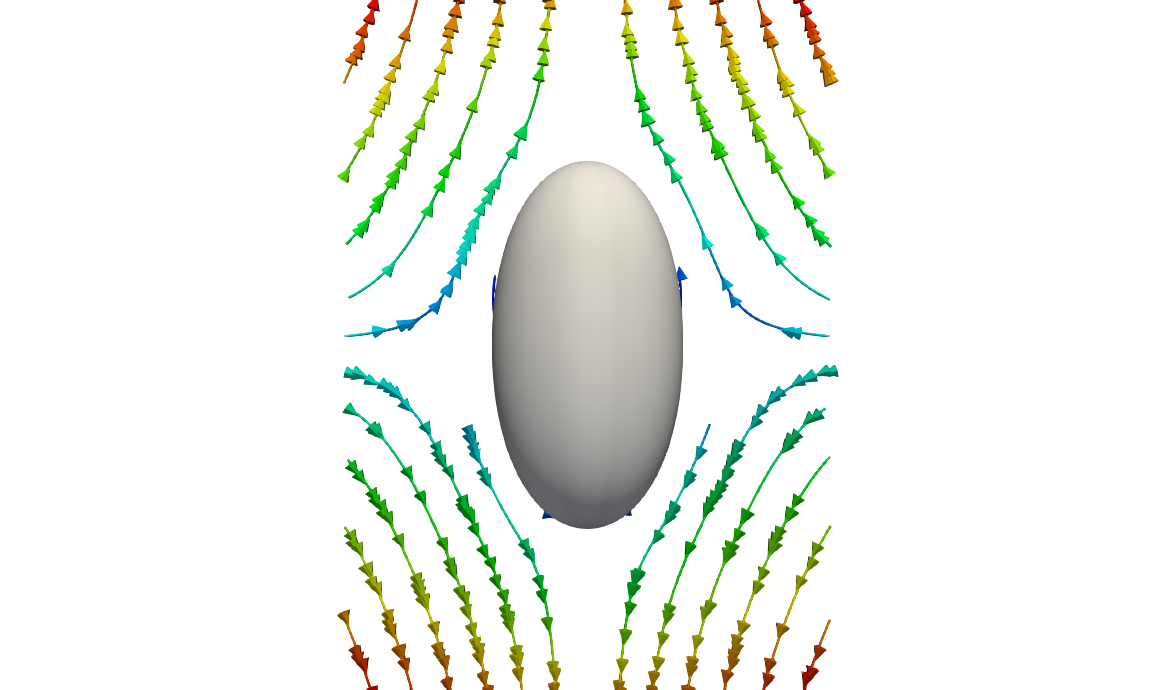}}
\caption{The flow layout of a droplet in an elongational (uniaxial extensional) flow. The image is a planar cut, with the flow being rotational symmetric around the elongated droplet axis in the image. The streamlines are coloured according to the velocity magnitude.}
\label{fig:droplet_hyperbolic}
\end{figure}

\begin{figure}[!htbp]
\centering
\includegraphics[scale=0.73]{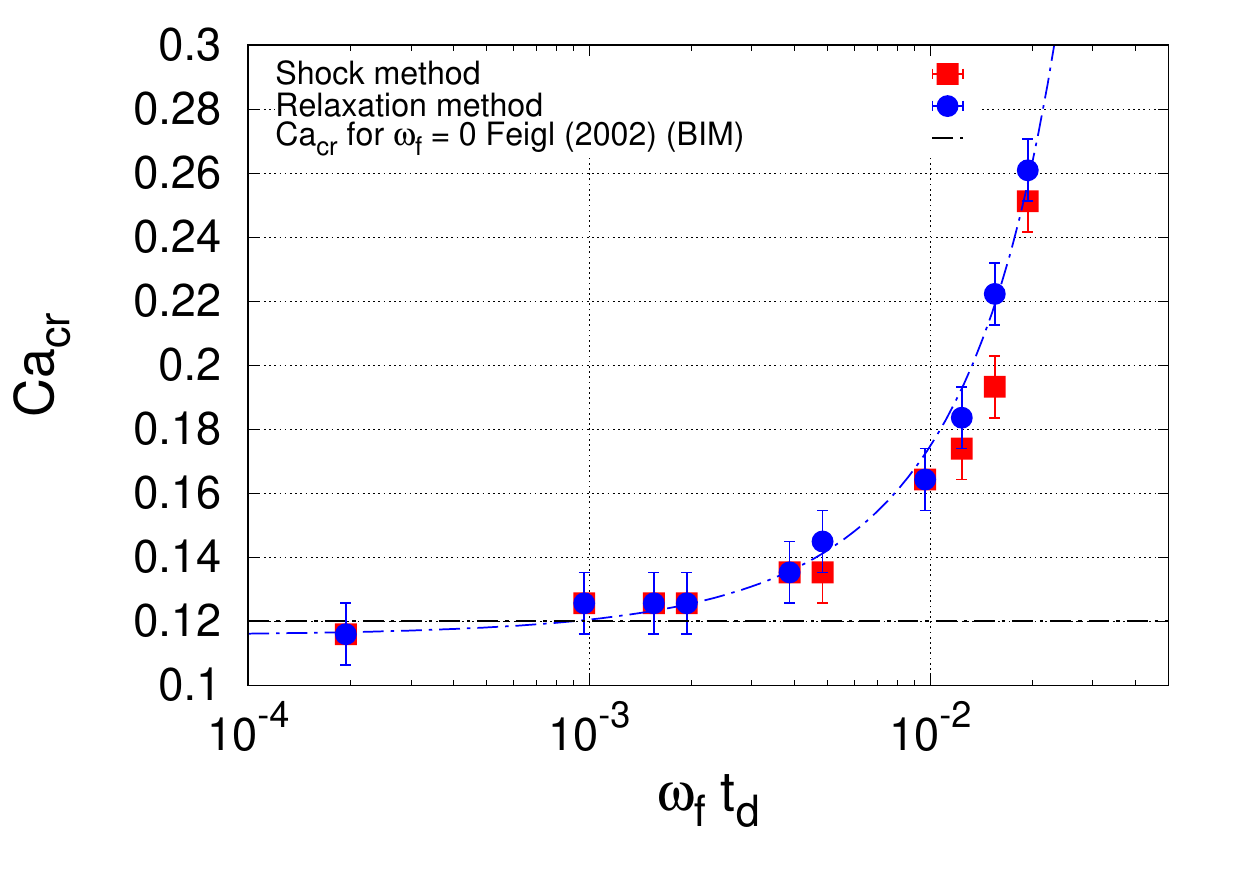}
\caption{Critical capillary number $\mbox{Ca}_{\text{cr}}$ against different frequencies $\omega_f t_d$ for a droplet in an unconfined elongational flow. We consider the two break up protocols, the \emph{shock method} and the \emph{relaxation method}. Even though the droplet break up is dependent on the shear rate frequency $\omega_f t_d$, a protocol mismatch does not occur, contrary to the case of the confined shear flow topology. The error bars are estimated via steps in the critical capillary number $\Delta \mbox{Ca}$.}\label{fig:elongational_critical}
\end{figure}

We now investigate the protocol mismatch in terms of the flow topology. Instead of an oscillatory shear flow, we consider break up in an elongational (or uniaxial extensional) flow; see Fig.~\ref{fig:droplet_hyperbolic}. This flow is by its very nature unconfined, so we would expect to not see a mismatch, as is the case for $\alpha = 0$ in the case of the confined shear flow; see Sec.~\ref{sec:confinement}. The results are shown in Fig.~\ref{fig:elongational_critical}. Interestingly, a mismatch between the two droplet protocols is absent and the predictions agree well with each other in terms of their respective errors. This shows that strong confinement ($\alpha \geq 0.75$) is necessary for the existence of the protcol mismatch shown in Fig.~\ref{fig:mm_critical}. Moreover, Fig.~\ref{fig:elongational_critical}, shows that droplet break up in an oscillatory elongational flow is frequency dependent, with an exponential dependence between the oscillation frequency $\omega_f t_d$ and the critical capillary number $\mbox{Ca}_\text{cr}$. The low-frequency limit matches the stationary flow predictions of~\cite{Feigl02}.

\section{Conclusions and Outlook}
\label{sec:conclusion}

We have shown that the interplay of varying start up conditions and strong confinement ratios can lead to qualitatively and quantitatively different  droplet break up conditions in stationary shear flows, unlike the stable equilibria found for varying start up conditions~\cite{Renardy08} or the ones found for varying degrees of confinement~\cite{Janssen10}. Having investigated the effects of inertia, confinement and flow topology, we conclude that the protocol mismatch between the shock and the relaxation method are due to a high degree of confinement for a droplet in a shear flow ($\alpha = 0.75$). However, the break up solution found via the relaxation method is only metastable, since it becomes unstable in the case of a time-dependent, oscillatory shear flow. The protocol mismatch is thus solely due to an extra metastable solution in a strongly confined shear flow and disappears in the presence of small perturbations (e.g., amplitude variations in an oscillatory shear flow) in accordance with the uniqueness of the Stokes solution~\cite{Janssen10,Renardy08}. We have also shown the dependency of the critical capillary number $\mbox{Ca}_{\text{cr}}$ on the normalized oscillation frequency $\omega_f t_d$ in both oscillatory shear and elongational flows. In the case of the elongational flow, $\mbox{Ca}_{\text{cr}}$ increases with increasing $\omega_f t_d$, whereas no simple functional dependence can be found for the oscillatory shear flow, since $\mbox{Ca}_{\text{cr}}$ also depends on the flow start up and degree of confinement. We should stress again that the results presented in this work are only valid for $\chi = 1$, since the viscosity ratio influences the breakup of a confined droplet~\cite{Vananroye06,Janssen08,Janssen10,VanPuyvelde08}. On the one hand, for small viscosity ratios $\chi \approx 0.3$ the confined shear flow stabilizes the droplet and breakup is more difficult to occur than for $\chi = 1$. On the other hand for large viscosity ratios $\chi \approx 5.0$ the confined droplet is destabilized and breakup is more likely to happen than for $\chi = 1$.~\cite{Janssen10,VanPuyvelde08}. It would be interesting to see whether the metastable solution can be found in an experimental setup or whether it is too prone to perturbations to manifests itself.

\section*{Acknowledgments}

The authors kindly acknowledge funding from the European Union's Framework Programme for Research and Innovation Horizon 2020 (2014 - 2020) under the Marie 
Sk\l{}odowska-Curie Grant Agreement No.642069 and funding from the European Research Council under the European Community's Seventh Framework Program, ERC Grant Agreement No 339032. The authors would also like to thank Fabio Bonaccorso, Dr Anupam Gupta, Dr Gianluca di Staso and Karun Datadien for their support.

\newpage



\bibliographystyle{spphys}
\bibliography{bibfile}

\end{document}